\documentclass[runningheads,a4paper]{llncs}
\usepackage{amsmath, amssymb, amsfonts}
\usepackage{subfigure}
\usepackage{graphicx}
\usepackage{setspace}
\usepackage{float}
\usepackage{lineno}
\usepackage{verbatim}

\newcommand{\Rspace}        {\mathbb {R}}

\usepackage{tikz}
\usetikzlibrary{calc}
\usetikzlibrary{patterns}
\tikzstyle{mybox} = [draw=black, fill=white,  thick,
    rectangle, inner sep=10pt, inner ysep=20pt]
\tikzstyle{mybox} = [draw=black, fill=white,  thick,
    rectangle, inner sep=2pt, inner ysep=2pt]

\title{Hardness of Segment Cover, Contiguous SAT and Visibility with Uncertain Obstacles\thanks{The work of Salman Parsa is supported by the National Science Foundation under Grant CCF-1614562 and funding from the Saint Louis University Research Institute.}}
\titlerunning{Hardness of Segment Cover}

\author{Sharareh Alipour\inst{1} \and Salman Parsa\inst{2}}
\institute{School of Computer Science, Institute for Research in Fundamental Sciences (IPM),{\tt alipour@ipm.ir}
\and
Computer Science Department, Saint Louis University, {\tt salman.parsa@slu.edu}}

\begin{document}
\thispagestyle{empty}
\maketitle

\begin{abstract}
We define the problem segment cover as follows. We are given a set of pairs of sub-intervals of the unit interval. The problem asks if there is a choice of a single interval from each pair such that the union of the chosen intervals covers the entire unit interval. This problem arises naturally while attempting to compute visibility between a point and a line segment in the plane in the presence of uncertain obstacles. Segment cover is equivalent to a restricted version of SAT which we call contiguous SAT. Consider a SAT with the following restrictions. An input formula is in CNF form and an ordering of the clauses is given in which clauses containing any fixed literal appear contiguously. We call this restricted problem contiguous SAT. Our main result is that the problems segment cover and contiguous SAT are NP-hard. We also discuss hardness of approximation for these problems.
\end{abstract}

\section{Introduction}\label{s:intro}
In this paper we consider two very related problems. One of them we call the segment cover problem and the other one contiguous SAT. These problems are encountered when trying to introduce a specific model of uncertainty into visibility problems, see \ref{s:motiv} below for the connection to this uncertain visibility model which has been the origin of this work. These two problems are very natural and we expect that they would be encountered in similar situations.

\subsection{Problem Statements and Results}
\begin{figure}[H]
  \centering
  \includegraphics[scale=.2]{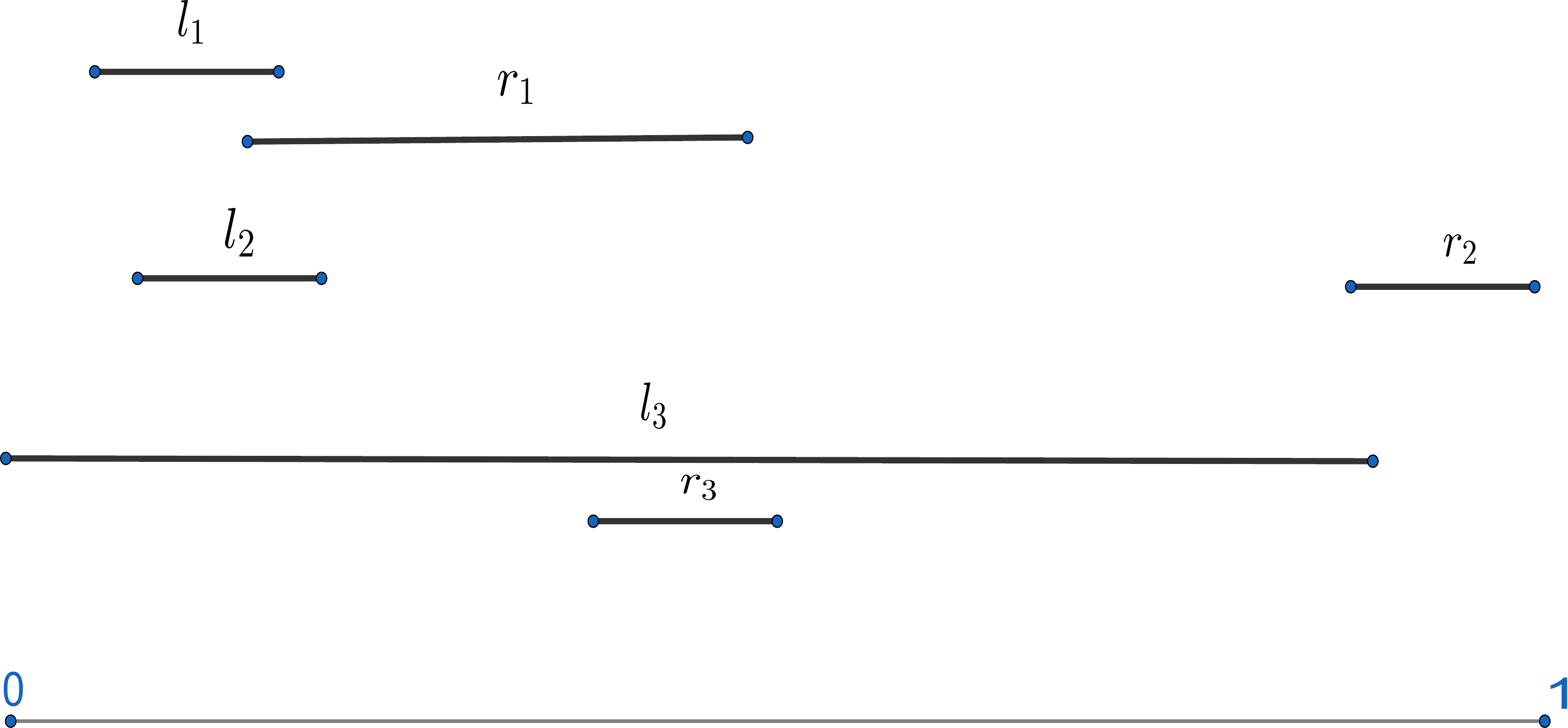}\\
  \caption{An instance of segment cover with 3 uncertain segments. The segments are depicted above the interval for clarity.}\label{fig:segmentcover}
\end{figure}

Our first problem is called the segment cover problem. Let $I$ be an interval of the real line, this interval is fixed once and for all and for simplicity we take $I=[0,1]$. We call a closed sub-interval of $I$ a \emph{segment}. An \emph{uncertain segment} is a pair $s=\{ l, r \}$ of two segments. An uncertain segment models the situation where we know that the ``real" segment is one of $l$ or $r$ but we do not know which one. Let $S=\{s_i, i=1,\ldots,n\}$ be a set of uncertain segments. The segment cover problem asks: Is there a choice of $l_i$ or $r_i$ (but not both) for each $i$ such that the union of the chosen segments is $I$. See Figure \ref{fig:segmentcover} for an example. In other words, if the uncertain segment $s_i$ is $l_i$ with probability $0 \leq p_i \leq 1$ and is $r_i$ with probability $1-p_i$, the problem asks to decide if the probability of the entire interval $I$ being covered is non-zero. We show in Section \ref{section:reduction} that the segment cover problem is NP-hard. We remark that if we require that any two segments (of all uncertain segments) are disjoint or coincide, then the NP-hardness result does not apply.  Therefore the NP-hardness is not immediate because of the presence of a choice. Also one is justified for having the first impression that this problem should not be NP-hard. We reduce 3SAT to segment cover. The method we use for the reduction is simple, however we believe it is novel.

Our second problem, the contiguous SAT is a restricted SAT problem. The input to contiguous SAT is i) a SAT instance in CNF form $C_1 \wedge C_2 \cdots \wedge C_m$, where the $C_i$'s are $m$ clauses, and ii) an ordering on the $C_i$. The instance must satisfy the \emph{contiguity} condition: any literal appears in a contiguous set of clauses with respect to the given ordering. It is convenient to assume that the subscripts of the $C_i$ respect the given ordering. Then the contiguity condition requires that for any positive literal $x$, the $i$ such that $x \in C_i$ form a contiguous set of numbers, and similarly for the literal $\neg x$. For example, the following formula is an input to contiguous SAT with the left-to-right ordering of the clauses.
$$ (x_1 \vee x_2 \vee x_3) \wedge ( x_1 \vee x_2 \vee \neg x_3) \wedge (x_1 \vee \neg x_3 \vee \neg x_1)$$
But the following is not a valid input.
$$ (x_1 \vee x_2 \vee x_3) \wedge ( \neg x_1 \vee x_2 \vee \neg x_3) \wedge (x_1 \vee \neg x_3 \vee \neg x_1)$$

Lemma \ref{l:equival} states that the above two problems are linear-time equivalent. Therefore, by the results of this paper, the contiguous SAT problem is also NP-hard. The segment cover problem provides a geometric view to contiguous SAT. This point of view has been useful in proving our results.

\paragraph{}
We also consider a special case of the segment cover problem in which the input segments are all of equal length. We call this problem \emph{all-equal segment cover}. We show in Theorem \ref{theorem:allequal} that all-equal segment cover is NP-hard. From this result follows that a problem called Best-Case Connectivity (BCU) in \cite{CEF-etal17,cham} is NP-hard even in dimension 1. Therefore we strengthen a main result of \cite{CEF-etal17,cham} considerably.

\begin{figure}[H]
  \centering
  \includegraphics[scale=0.2]{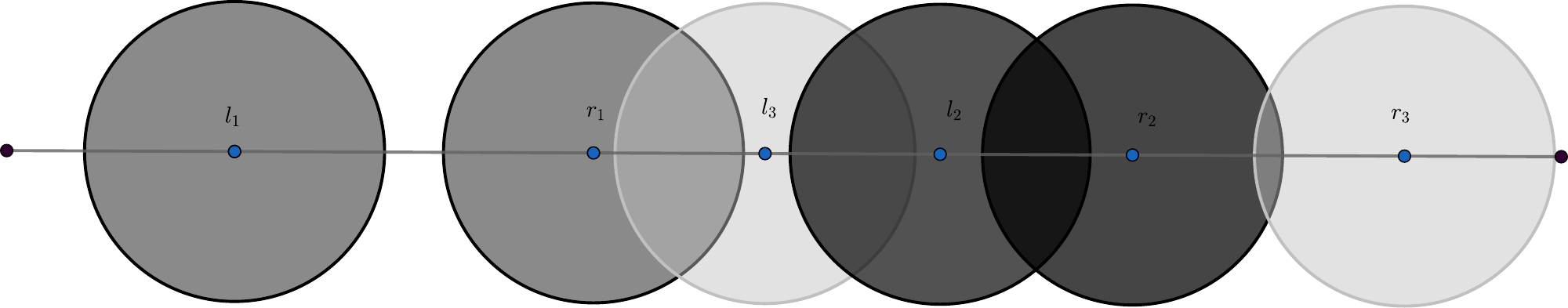}\\
  \caption{What is the minimum radius such that the union of three balls, one chosen from each color, is connected?}\label{fig:bcu}
\end{figure}

The BCU problem is defined as follows.  Let $R_1, \ldots, R_n$ be $n$ closed regions in the $d$-dimensional Euclidean space. These are called \emph{uncertainty regions}. Find the minimum value of $r$ satisfying: there exist points $p_i \in R_i$ such that the union of the balls of radius $r$ centered at the $p_i$, $\bigcup_i B(p_i,r)$, is connected.
In \cite{CEF-etal17,cham} it is shown that this problem is NP-hard in the plane, $d=2$, even when each uncertainty region is a pair of points. They leave the case $d=1$ open. We reduce this problem to the all-equal segment cover in Corollary \ref{c:BCU}. See Figure \ref{fig:bcu} for an example.

In Section \ref{s:approx} we also prove hardness of approximation results for the segment cover problem.

\subsection{Relations to Uncertain Visibility }\label{s:motiv}
This section describes the motivation behind our problems and can be skipped.
One of the basic problems in computational geometry is computing visibility in various configurations of points and obstacles in the Euclidean plane. Visibility plays an important role in robotics and computer graphics, among other areas. In robotics, for example, the efficient exploration of an unknown environment requires computing the visibility polygon of the robot or test whether the robot sees a specific object or not.

Suppose that we are given a set $S$ of $n$ obstacles in the plane, say in the form of convex polygons. Two points $p$ and $q$ are \emph{visible} to each other if their connecting line segment does not intersect any of the obstacles in $S$. A line segment $t \subset \Rspace^2$ is visible to a point $p$ if $p$ is visible to at least one point of the line segment $t$.

In a version of \emph{visibility testing problem} we are given a set $S$ of $n$ obstacles in $\mathbb{R}^2$ and our goal is to preprocess $S$ so that we can quickly answer a query of the form: is a query segment $t \subset \mathbb{R}^2$ visible from a query point $q \in \mathbb{R}^2$? For instance $q$ is a camera and the segment represents an object of interest.

\begin{figure}[H]
  \centering
  \includegraphics[scale=0.2]{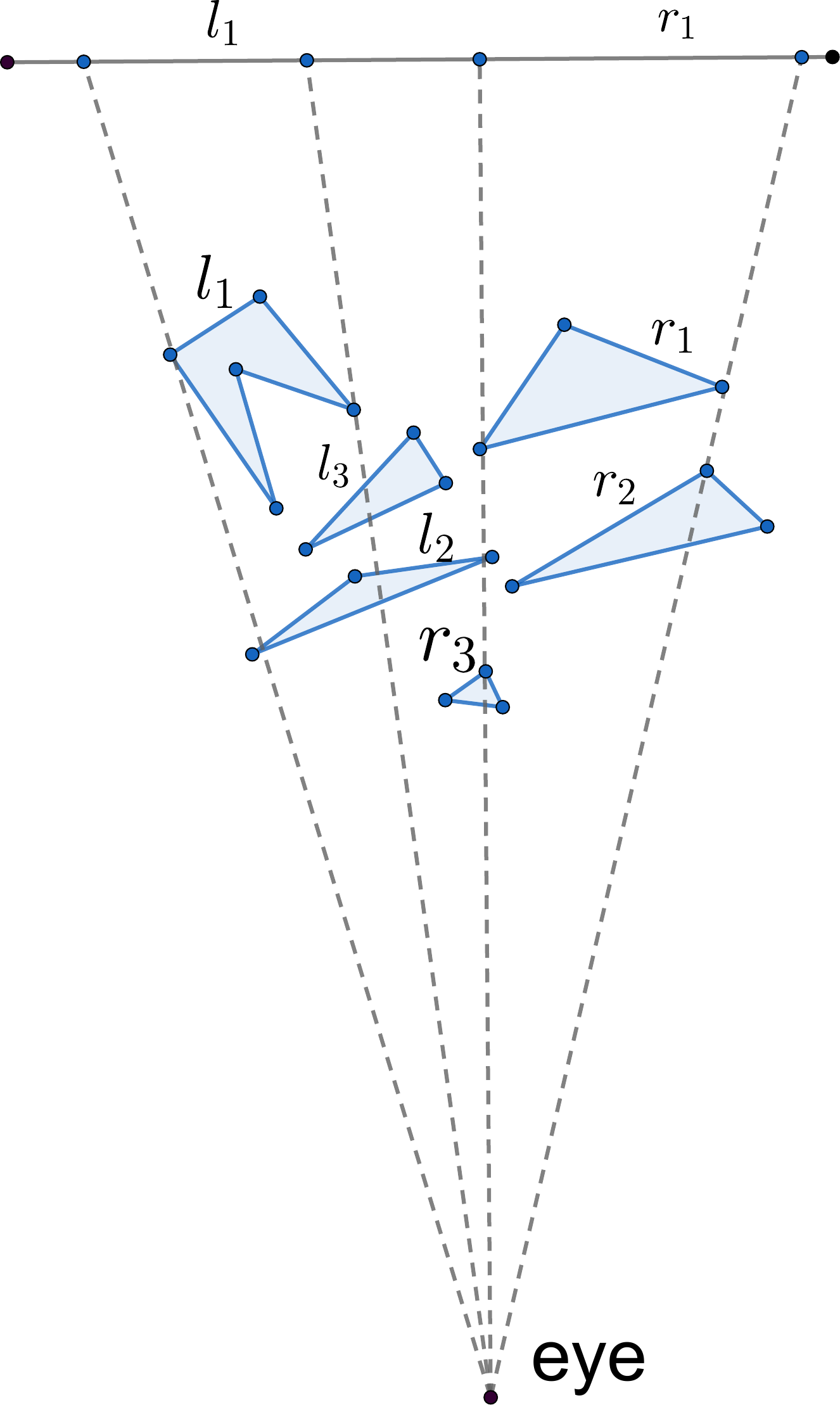}\\
  \caption{3 uncertain obstacles, each with two possibilities $l_i$,$r_i$, the projection each uncertain obstacle gives an uncertain segment, one such uncertain segment is depicted  }\label{fig:uncertainvis}
\end{figure}

Recently, there has been some attention to situations wherein there is uncertainty in the obstacle positions. Then, two points are visible to each other with a certain probability. Examples of these kind of problem can be found in \cite{BKLS15,AAGM17}. We introduce uncertainty into the set of obstacles $S$ as follows. Each obstacle exists in one of two possible locations with given probabilities. Consequently, the set $S$ is replaced by a set of pairs $\{ l_i,r_i\}$ of obstacles. We denote the set of pairs again by $S$. The \emph{probabilistic visibility testing problem} asks to preprocess $S$ to answer the following query: What is the probability that a given segment $I \subset \mathbb{R}^2$ be visible from a given query point $q$? It was shown in \cite{AAGM17} that computing this probability exactly is $\#P$-hard. Note that for a fixed $q$, each possible location of an obstacle covers a sub-segment of $I$. Then the the probability that the interval $I$  be visible to $q$ is 1 if and only if, for any choices of $l_i$ or $r_i$ the projections of the chosen obstacles to $I$ do not cover the interval $I$. This is our segment cover problem. See Figure \ref{fig:uncertainvis}. Our results imply that deciding that a given point $q$ can always see a given segment $I$ is NP-hard, in the presence of uncertain obstacles.

\subsection{Related Work}\label{subsection:relatedwork}

The problems around the visibility concept have been an active research area since the beginning of computational geometry. Point and edge visibility\cite{AvTo81,BHKM04,GHLST87,ChGu89}, the art gallery problem~\cite{ORou87}, the watchman route problem\cite{CaJoNi99,Mitch11}, visibility graphs and their recognition\cite{Gho07,GuMo10} are among the topics of interest in this field.  Uncertainty is very natural in applications and indeed has been studied from a more practical viewpoint, like robot motion planning \cite{MoMiSh99,Bri92,Don90,MCGBTo02,CoPe08}.

There are many restrictions of SAT that are NP-complete. They include $k$-SAT, $k\geq3$, NAE-SAT,  1-in-3 SAT~\cite{Scha78}, planar 3-SAT~\cite{Lich82}, planar 1-in-3 SAT~\cite{Laro93}, monotone planar cubic 1-in-3 SAT~\cite{MoRo01}, 4-bounded planar 3-connected 3-SAT~\cite{Kra94}.
Let us denote by $(r,s)$-SAT the SAT problem restricted to clauses containing exactly $r$ variables, and each variable appearing in at most $s$ clauses. Tovey~\cite{Tov84} has shown that $(3,3)$-SAT is always satisfiable and $(3,4)$-SAT is NP-Complete. In addition, it is proved in \cite{PaYa91} that 3-SAT restricted to instances where each variable appears at most three times is NP-Complete. A stronger result, proved in Dahlhaus et al.~\cite{DJPSY94}, states that planar 3-SAT in which each variables appears exactly three times, and twice with one literal, a third time as the other literal, is still NP-Complete. These results will be used in Section~\ref{section:reduction}. We make use of the hardness of SAT instances where each variable is restricted to a few clauses in our proof of hardness of all-equal segment cover. None of these special cases directly imply hardness of contiguous SAT. Indeed although our reduction is quite simple its method is novel to the best of our knowledge.

\section{Reduction}\label{section:reduction}

In this section we reduce 3SAT to segment cover.

\paragraph{Convention}We make the following convention that by a \emph{literal} of a SAT instance we mean
an appearance of $x$ or $\neg x$ for some variable $x$ in $\phi$. Therefore, a literal is uniquely determined by determining its clause and
a number. In addition, if an interval is the union of sub-intervals such that the sub-intervals share only endpoints with each other and are otherwise disjoint, by abuse of notation and for simplicity, we say that the interval is a \emph{disjoint union} of the sub-intervals.

\paragraph*{}
We begin by observing the following.
\begin{lemma}\label{l:equival}
The problems segment cover and contiguous SAT are equivalent with linear-time reductions.
\end{lemma}

\begin{figure}[H]
  \centering
  \includegraphics[scale=0.2]{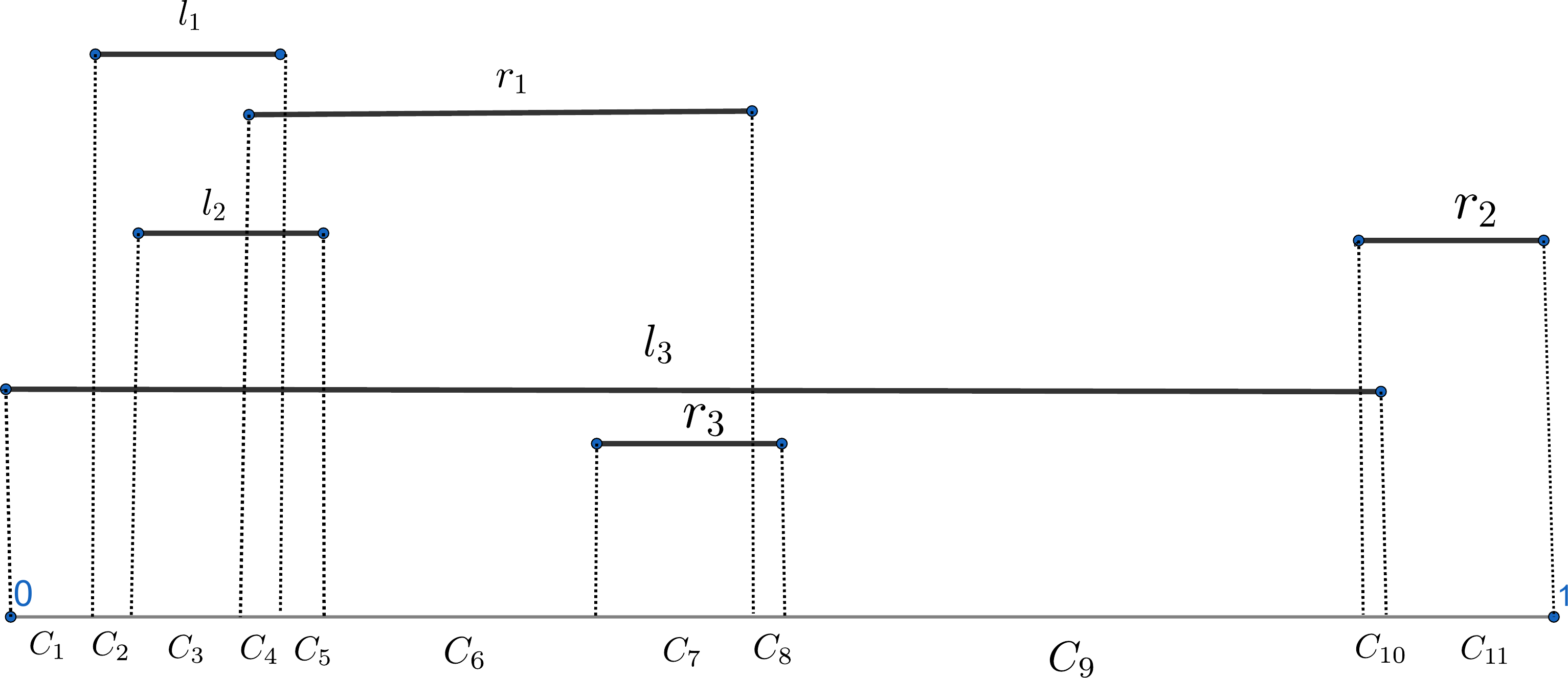}\\
  \caption{defining a SAT instance from a segment cover instance, for example $C_5=(\neg x_1 \vee x_2 \vee  x_3 )$}\label{fig:lemma}
\end{figure}

\begin{proof}
Given an instant of segment cover construct an instance of contiguous SAT as follows. Partition the interval $I$
using the endpoints of all of the given segments into subintervals $J_i$. The $J_i$ are closed sub-intervals
that only share (one or two) endpoints with the neighboring sub-intervals and otherwise are disjoint. For each sub-interval $J_i$ we
define a clause $C_i$. For any uncertain segment $s_j=\{ l_j, r_j \}$ define a variable $x_j$. For all $i$ and $j$, add the literal $x_j$
to the $C_i$ if $J_i$ is covered by $l_j$ and add the literal $\neg x_j$ to the $C_i$ if $J_i$ is covered by $r_j$. Order the
$C_i$ using the left-to-right ordering of the sub-intervals $J_i$. This defines an instance of contiguous SAT. See Figure \ref{fig:lemma}.

If the contiguous SAT instance we constructed is satisfiable then we choose for $s_j$, the segment $l_j$ if $x_j=1$ and $r_j$ if $x_j=0$.
These choices cover all of $I$. If the segment cover is satisfiable then the contiguous SAT instance is satisfiable.
Therefore segment cover reduces to contiguous SAT.

Similarly given an instance of contiguous SAT with $m$ clauses one can construct an instance
of segment cover by i) partitioning the segment $I$ into $m$ sub-intervals $J_i=[(i-1)/m,i/m]$, $i=1, \cdots,m$, and associating $J_i$ to $C_i$ and ii) defining a segment $s_j$ for the variable $x_j$ and defining $l_j$ and $r_j$ by concatenating the sub-intervals corresponding to clauses in which $x_j$ or $\neg x_j$ appear. We omit the details.\qed
\end{proof}

We now start our reduction of 3SAT to segment cover. More over we assume that clauses contain exactly three literals. Let $\phi$ be the given 3SAT formula with $s$ clauses, $C_1, \ldots, C_s$. For simplicity of presentation assume each variable appears at least once as a positive literal and at least once as a negative literal in $\phi$. We first divide the interval $I$ into $s$ disjoint sub-intervals $B_j = [(j-1)/s, j/s]$, $j=1, \ldots,s$. Next, we define an arbitrary one-one correspondence between the clauses and the intervals $B_j$. For simplicity we take $B_j$ to be the sub-interval corresponding to $C_j$.

\begin{figure}[H]
  \centering
  \includegraphics[scale=.2]{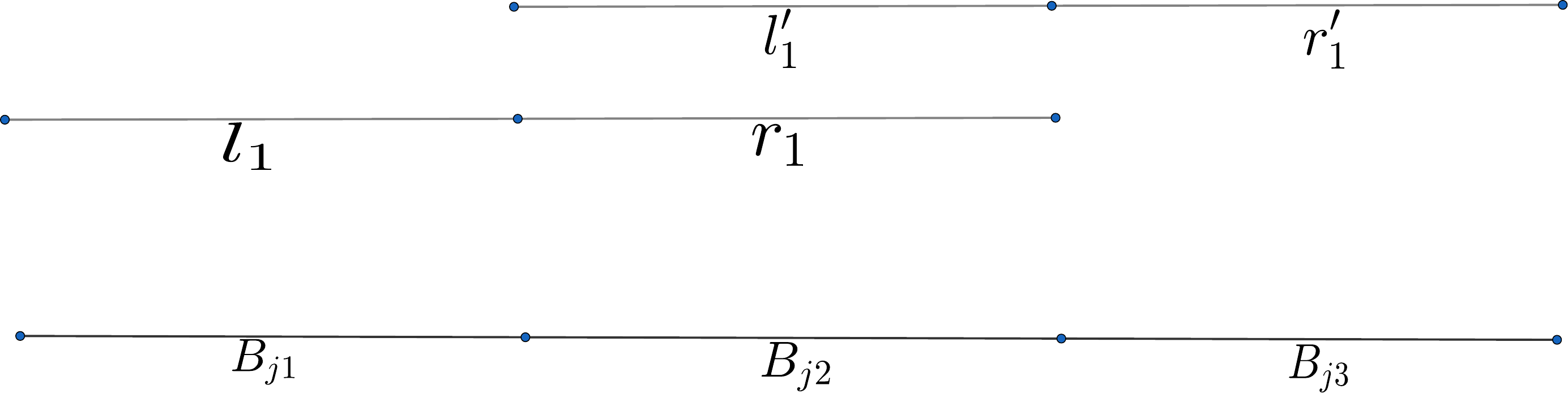}\\
  \caption{the uncertain segments defining $T_j$, the segment are shown above the interval for visual purposes}\label{fig:clausesegments}
\end{figure}

\paragraph*{Clause uncertain segments}
We partition each  $B_j$ into three equal parts, $B_{j1}$, $B_{j2}$ and $B_{j3}$. Next we consider the set $T_j = \{\{B_{j1},B_{j2}\}, \{B_{j2},B_{j3}\}   \}$ containing the two uncertain segments $\{B_{j1},B_{j2}\}$ and $\{B_{j2},B_{j3}\}$ as shown in Figure~\ref{fig:clausesegments}. $T_j$ has the following property.

\begin{lemma}
For any choice of segments from the uncertain segments of $T_j$, at most two intervals among $B_{j1},B_{j2}$ and $B_{j3}$ are covered.
\end{lemma}

We define an arbitrary one-one correspondence between the literals of $C_j$ and the sub-intervals $B_{j1}$, $B_{j2}$ and $B_{j3}$. We denote this correspondence by $\alpha_j: L_j(\phi) \rightarrow \{B_{j1},B_{j2},B_{j3}\}$, where $L_j(\phi)$ denotes the set of of literals of $C_j$. We denote by $\alpha$ the one-one correspondence between all appearances of literals in $\phi$ and all $B_{jk}$, $j=1, \ldots,s$, $k=1,2,3$ which is defined by the $\alpha_j$. Again for simplicity we can take $\alpha$ to be the correspondence suggested by the subscripts, that is if $C_j=(\lambda_1 \vee \lambda_2 \vee \lambda_3 )$ then $\alpha(\lambda_i) = B_{ji}$, $i=1,2,3$.

\paragraph*{Variable uncertain segments}
Let $x_1, \ldots, x_m$ be the variables of the given formula $\phi$. We shall construct a collection of uncertain intervals $S_i$ for the variable $x_i$ . For each variable, these uncertain intervals are defined by means of a complete bipartite graph denoted $G_i$, $i= 1, \ldots, m$. The vertices of $G_i$ are the literals of $\phi$. The vertices are divided into two parts denoted $P_i$ and $N_i$, namely positive and negative literals. This finishes the definition of $G_i$.

For each $i$, using $G_i$, we define the set $S_i$ as follows. Let $v \in P_i \cup N_i$ be a literals of $x_i$ and take $J=\alpha(v)$. Let $d=d(v)$ be the degree of $v$ in the graph $G_i$. Partition the interval $J$ into $d$ disjoint (sharing only endpoints) sub-intervals, and define an arbitrary one-one correspondence $\beta_v$ between the edges incident on $v$ and these sub-intervals. Perform this subdivision for the intervals $\alpha(v)$ for all $v \in P_i \cup N_i$. We obtain thus a set of one-one correspondences $\beta_v$ between the edges incident to the vertex $v$ and subintervals of $J=\alpha(v)$.

Now the uncertain segments $S_i$ are defined by the edges of the graph $G_i$ and their corresponding sub-intervals. In more detail, let $e = \{ v_P,v_N \}, v_P \in P_i, v_N \in N_i$ be an edge of the graph $G_i$. Then $e$ determines the segment $\beta_{v_P}(e)$ inside $\alpha(v_P)$, and the segment $\beta_{v_N}(e)$ inside $\alpha(v_N)$. Then $s_e \in S_i$ is defined as the uncertain segment containing these two segments.

The segment cover instance for $\phi$ is the set of uncertain segments $S = \bigcup_{i=1}^{m} S_i \cup \bigcup_{i=1}^{s} T_i$.

\subsection{Correctness of the Reduction}
In this section we show that there is a covering of the unit interval with the uncertain segments $S$ if and only if the given sentence $\phi$ is satisfiable.

Assume that $\phi$ is satisfiable. Observe that each uncertain segment $s \in S_i \subset S$, has a \emph{positive segment} and a \emph{negative segment}. Namely, the positive segment is the one which corresponds to the incidence of the edge to a vertex in the positive part of $G_i$, i.e., $P_i$, and analogously for the negative segment. Hence we can write $s= \{s_p, s_n\}$ where $s_p=\beta_{v_P}(e), s_n=\beta_{v_N}(e)$, where $e=\{v_P,v_N\}$ is the edge defining $s$. Now assume $x_i$ takes the value 1 (=true) in the assignment that satisfies $\phi$. We choose $s_p$, otherwise we choose $s_n$, for all $s \in S_i$.

We spell out the following fact.
\begin{lemma}\label{lemma:subintervalpart}
An interval $B_{j_0k_0}$, for some $j_0 \in \{1,\ldots,s\}$, $k_0\in \{1,2,3\}$, satisfies $\alpha(v)= B_{j_0k_0}$ for some $v \in P_i$ if and only if $x_i$ has a positive literal in $C_{j_0}$. Analogously, an interval $B_{j_0k_0}$ satisfies $\alpha(v)= B_{j_0k_0}$ for some  $v \in N_i$ if and only if $x_i$ has a negative literal in $C_{j_0}$.
\end{lemma}

From the above lemma, whenever we choose the uncertain segments as above, since each clause is satisfied, each clause $C_j$ has a literal $v$ (=vertex in some graph) all of whose incident edges have chosen that vertex. Hence, $\alpha(v)$ among $B_{j1},B_{j2}$ and $B_{j3}$ is covered. It remains to cover the two remaining intervals. This is easily done by a suitable choice for the uncertain segments of the set $T_j$. This finishes one direction of the proof.

Consider now the other direction. We have to show that if there is a choice for each uncertain segment $s \in S$, such that the unit interval is covered, then, there is an assignment of 0 and 1 to the variables $x_i$ that satisfies the given formula $\phi$. Consider a clause $C_j = (\lambda_1 \vee \lambda_2 \vee \lambda_3)$, where $\lambda_i$ are literals. And let $x_{i_1}, x_{i_2},x_{i_3}$ be the corresponding variables. The interval $B_j$ is covered by the chosen segments. Recall that the uncertain segments correspond to the edges of the graphs $G_i$ (other than elements of the $T_j$) and that a choice of an interval for an uncertain segment is equivalent to choosing one endpoint of the corresponding edge.

\begin{lemma}\label{lemma:literalchoice}
Assume there is a choice of uncertain segments such that $I$ is covered. Consider the graphs $G_{i_1}, G_{i_2}, G_{i_3}$ of variables of any clause $C_j = (\lambda_1 \vee \lambda_2 \vee \lambda_3)$. There exists at least one vertex $\lambda$ among $\lambda_i$, $i=1,2,3$, such that, each edge incident on $\lambda$ has chosen $\lambda$.
\end{lemma}

\begin{proof}
The uncertain segments in $T_j$ leave at least one of $B_{j1},B_{j2}$ and $B_{j3}$ uncovered, wlog, let it be $B_{j1}$.
The interval $B_{j1}$ has to be covered using the uncertain segments of $S$. All the edges incident on $\lambda_1$ are required to choose $\lambda_1$, otherwise, some part of the interval $B_{j1}=\alpha(\lambda_1)$ would remain uncovered.\qed
\end{proof}

To construct an assignment from the choices of uncertain segments $S$ we do as follows. Consider any clause $B_j$. Lemma \ref{lemma:literalchoice} gives a $k \in \{1,2,3 \}$ such that a vertex in $G_{i_k}$ is chosen by all its incident edges. If the vertex is in $P_{i_k}$, set $x_{i_k} = 1$, otherwise set $x_{i_k}$ to be 0.

First, we prove our assignment is well-defined. Assume for the contrary that $x_i$ has been assigned both 0 and 1.
Let $v_p \in P_i$ be the vertex based on which we have assigned 1 to $x_i$, and let $v_n \in N_i$ be the vertex based on which
we have assigned 0 to $x_i$. Therefore all of the edges incident to $v_p$ have chosen $v_p$ and all the edges incident on $v_n$ have chosen
$v_n$. But this is a contradiction since $G_i$ is a complete bi-partite graph.

Second, we show that the assignment satisfies all of the clauses. Take a clause $C_j$ and consider its corresponding interval $B_j$. Let $x$ be the vertex returned by Lemma \ref{lemma:literalchoice} and $v$ the vertex of the graph of $x$ all of whose incident edges have chosen it. If $v$ is a positive literal in $C_j$, $v$ is in the positive part of $G$. Hence our procedure setting $x=1$ satisfies the clause. If $x$ has a negative literal in $C_j$, then $v$ appears in the negative part of $G$. Hence setting $x=0$ will satisfy $C_j$. This finishes the proof of the correctness of the reduction.

We now bound the run-time of the reduction procedure. Assume the given formula $\phi$ is an arbitrary 3SAT instance. In linear time in number of clauses we construct the sets $T_j$ of uncertain segments. Let variable $x_i$ appear in $p_i$ clauses as a positive literal and in $n_i$ clauses as a negative literal. Then the graph $G_i$ is $K_{p_i,n_i}$ and has $p_in_i$ edges. Thus our reduction is of complexity $O(s + \sum_{i=1}^m p_in_i)$.

\begin{theorem}\label{theorem:main}
The problems segment cover and contiguous SAT are NP-Complete.
\end{theorem}

\paragraph{Remark} If we start by the NP-Complete problem studied by \cite{DJPSY94} in which each variable appears at most three times  then the number of our uncertain segments is exactly $2s+2m$ (this would require also dealing with clauses with two literals).

\section{All-Equal Segment Cover}\label{section:allequal}
In this section we strengthen our result to show that the segment cover remains NP-complete even when we require that the lengths of the intervals all be equal. We call this problem all-equal segment cover. We will later deduce that the problem BCU of \cite{CEF-etal17,cham} (defined in Section \ref{s:intro}) is also NP-complete for $d=1$.

\begin{figure}
  \centering
  \includegraphics[scale=0.25]{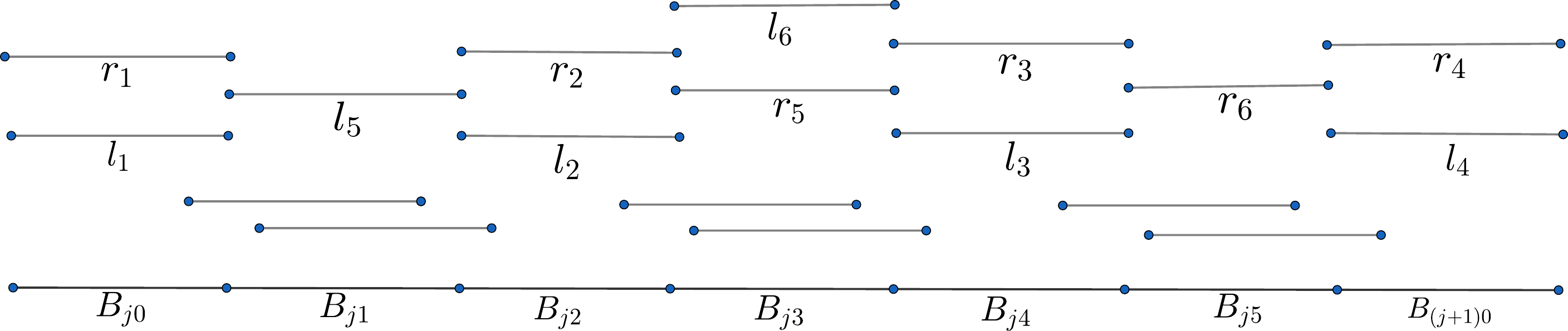}\\
  \caption{The labelled intervals on the top define uncertain segments $T_j$, unlabelled ones in the bottom define sets $S_i$.}\label{fig:allequal}
\end{figure}

We now describe the modifications to the reduction necessary to keep all the intervals the same length. Observe that we can make sure that the intervals $B_{j1}, B_{j2}, B_{j3}$, for all $j$, have equal length. It remains to make sure that the intervals in the uncertain segments from the $S_i$ have equal length and have also length equal to the $B_{ji}$. For simplicity in this argument, we will start by a special 3SAT problem, namely, the one considered by \cite{DJPSY94}. They have proved that planar 3-SAT remains NP-Complete when each variable appears at most three times, once as one literal, twice as the other. If a clause contains only two literals we change the intervals $T_j$ of Figure \ref{fig:clausesegments} accordingly. When applying our reduction to this type of formulas, we will see that in the final uncertain segments intervals $B_{j1}, B_{j2},B_{j3}$ are divided into at most two smaller intervals. With these preliminaries in mind, we will substitute the intervals in the Figure \ref{fig:allequal} for the corresponding intervals from the construction of Section \ref{section:reduction}. In this figure $B_{j1}, B_{j3}, B_{j5}$ play the roles of $B_{j1}, B_{j2}$ and $B_{j3}$ of the original reduction.
Note that we have assumed in the figure that the worst case happens, i.e., each three of the sub-intervals is divided. The other cases are simpler.

\begin{theorem}\label{theorem:allequal}
The problem all-equal segment cover is NP-Complete.
\end{theorem}
\begin{proof}
Consider a clause $C_j$ and its sub-interval $B_j$. To the set $T_j$ of the original reduction we add $s-1$ new uncertain segments, each of them consisting of two copies of the same segment. This insures that certain subsets of the interval $I$ are always covered, see Figure \ref{fig:allequal} (for instance, $\{ s_1,s'_1\}$ is such an uncertain segment). The set $S_i$ of uncertain segments for the variable $x_i$ is defined just as in the original reduction, but with the modification that a vertex interval $B_{jk}$ is not partitioned, rather the sub-intervals for the at most two incident edges are copies of the interval $B_{jk}$, one of them slightly moved to the right, the other slightly moved to the left.

We need to check that the new intervals have the required properties used in the reduction.
As before, at most two of the intervals $B_{j1}, B_{j3}$ and $B_{j5}$ can be covered by the uncertain segments from (updated) $T_j$. It is easily checked that any interval, say $B_{j1}$, which is not covered, can only be covered when both of the intervals of the incident edges are present. Hence, the same correctness argument applies here as well.\qed
\end{proof}

We next show that the optimization problem called BCU and studied in \cite{CEF-etal17,cham} is NP-complete on the real line. For the definition of BCU refer to Section \ref{s:intro}.
\begin{corollary}\label{c:BCU}
The 1-dimensional BCU is NP-complete.
\end{corollary}
\begin{proof}
Let the set $\{s_1, \ldots,s_n \}$ of uncertain segments be an instance of all-equal segment cover. For each $s_i$, construct an uncertain region $u_i$ containing two points, namely, the midpoints of the two intervals in $s_i$. We add two more regions defined as follows. Let $x_l$ be the smallest coordinate and $x_r$ be the largest coordinate of any midpoint. Moreover, let $r$ be half the length of an interval. Define $u_0 = \{ x_l-2r, x_l-3r \}$ and $u_{n+1} =\{x_r+2r, x_r + 3r \}$. Add these two sets to the problem instance. Then the $u_i$ define an instance of BCU. An algorithm solving BCU returns a minimum $r'$ such that there are $n+2$ disks of radius $r'$, with centers at the points of the $u_i$, one center from each $u_i$, such that the area they cover is connected. Because of $u_0, u_{n+1}$ we have always $r' \geq r$. Moreover, $r'=r$ if and only if the answer to the original all-equal segment cover is affirmative.

\end{proof}

\section{Approximation}\label{s:approx}
In this section we consider the approximation of the segment cover problem.  We can define two natural approximation problems. The first, called \emph{max-segment cover}, or \emph{max-SC} for short, asks to choose one interval from each uncertain segment such that the union of the resulting intervals is of maximum length possible among all the choices. Here we have extended the meaning of length of an interval to the length of a union of disjoint intervals
in the obvious way. Therefor, max-SC asks for maximum coverage.

The second, called \emph{contiguous max-SC}, requires a choice of an interval from each uncertain segment such that a maximum-length connected interval is obtained, among all connected intervals of all choices. Therefore contiguous max-SC asks for maximum connected coverage. In this section we discuss the approximation problem max-sc and leave more specialized study of contiguous max-SC as an open problem.

\subsection{Hardness of Approximation for Max-SC}
We first prove hardness of approximation for max-SC. Let max-E3SAT be max-3SAT restricted to formulas in which each clause contains exactly three literals.

\begin{theorem}\label{theorem:maxsc}
Let $c'$ be a ratio beyond which it is NP-hard to approximate max-E3SAT. Then it is NP-hard to approximate max-SC with ratio larger than $c = \frac{c'+2}{3}$.
\end{theorem}

\begin{proof}
Suppose we are given an instance $\phi$ of max-E3SAT with $n$ variables. We shall apply the reduction of Section \ref{section:reduction} to $\phi$ and obtain an instance of max-SC, however, we need some modifications. Consider a graph $G_i$ constructed in the reduction. If $|P_i|=|N_i|$ we leave the graph as it is, otherwise, let $|P_i| < |N_i|$. We add $|N_i|-|P_i|$ dummy vertices to $|P_i|$ to make the two sets equal. We do analogously in the other case. Let $\tilde{G}_i$ denote the modified graphs, $i=1,\ldots,n$. We build the uncertain segments $S_i$ from $\tilde{G}_i$ as follows. Let $v$ be a vertex of $\tilde{G}_i$ and $m=|\tilde{P}_i|=|\tilde{N}_i|$. If $v$ is not a dummy vertex, it has associated with it a sub-interval of a clause-interval. We make sure all these sub-intervals have length 1, and a clause interval has length 3. If $v$ is a dummy vertex, associate to it the fixed interval $J'$ of very small length $\epsilon>0$, anywhere outside all of the clause intervals.

Next, we build the uncertain segments $S_i$ as before from the graphs $\tilde{G}_i$ and associated intervals.
Let $W$ be the total length of the union of the intervals of uncertain segments $S_i$, then by construction
$$W = 3s+\epsilon.$$
Note that any two intervals of (possibly different) uncertain segments defined here are disjoint other than when both intervals are sub-intervals of $J'$.

We run the approximation algorithm for max-SC on our instance. The algorithm makes a choice from each uncertain segment. We modify this choice slightly. If any uncertain segment has chosen a sub-interval of $J'$ we reverse this choice. It is clear that at the end we have at worst decreased the total approximated length by $\epsilon$. And we have not decreased the approximated length over the original clause intervals.

Observe that the total length that the uncertain segments chosen from $T_i$ contribute is at most $2W/3=s$. If from any clause-interval the choice from $T_i$ covers only $1/3$ of the interval, then the middle interval $B_{j2}$ is covered.
We change the choices so that only $1/3$ of the interval is not covered, by covering either of $B_{j1}$ or $B_{j3}$. This insures that from any clause interval exactly one sub-interval is not covered by the $T_i$.

Now from the modified choice of uncertain segments and the graphs $G_i$ define the graphs $G'_i$ as follows.
For each $i$, from the graph $\tilde{G}_i$, remove any vertex whose interval is covered in the approximation by intervals from $T_i$. Denote the new graph by $G'_i$. The total length of the intervals corresponding to the non-dummy vertices of $G'_i$ is $W/3=s$. Next, define an assignment as follows. We distinguish five cases from each other.

\begin{itemize}
\item Case 1: The graph has original vertices in positive part only, and, dummy vertices are in positive part. For any edge $e \in \tilde{G}_i$ that is not incident with a dummy vertex, we redirect the choice to the positive side. Note that since any interval we uncover is covered by $T_i$ this does not decrease the length of the approximation. After these re-directions, any non-dummy vertex in the positive side of $\tilde{G}_i$ has all its sub-intervals chosen.

\item Case 2: The graph has original vertices in positive part only, and, dummy vertices are in negative part. For any edge $e \in \tilde{G}_i$ that is not incident with a dummy vertex, we redirect the choice to the positive side. Recall that all the other edges have also chosen the positive side. Then again after this re-direction of choice all the vertices in positive part of $\tilde{G}_i$, have their intervals covered. Again this operation does not decrease the total approximated length.

\item Cases 3,4: These are analogous to the previous cases, where non-dummy vertices appear in the negative part only. We perform analogously as in those cases.

\item Case 5: The graph has non-dummy vertices in both parts, or it has only dummy vertices. In this case, we can assign an arbitrary value to $x_i$. We choose the side which does not have dummy vertices and redirect all the edges of $\tilde{G}_i$ towards that side. Re-direction of the choice for an edge not incident on a dummy vertex does not change the approximated weight. Also we had set the choice for edges incident on dummy vertices away from them. It follows that all the intervals associated to the vertices of the chosen side are covered.

\end{itemize}

Thus we have defined an assignment. Now we compute the number of clauses satisfied by our assignment. The length not covered by the $T_i$ and covered by the $S_i$ in the approximation is at least $c(W+\epsilon)-\frac{2}{3}W$. After the above redirection of the choices, an interval corresponding to a clause is either all covered or covered in exactly $2/3$ of its length. Therefore, $c(W+\epsilon)-\frac{2}{3}W$ is (lower bound for) the total number of the intervals satisfied by our assignment. For any algorithm that runs in polynomial times we must have $c(W+\epsilon)-\frac{2}{3}W = 3cs+c\epsilon-2s<c's$.
This implies $$c<\frac{c'+2}{3+\frac{\epsilon}{s}}.$$
The claim follows.\qed
\end{proof}

\paragraph*{Remark} By a seminal result of H{\aa}stad~\cite{Has01} max-E3SAT cannot be approximated by a ratio larger than $7/8$. Using this result the above theorem implies that max-SC cannot be approximated beyond the ration $23/24$, unless P=NP.

\paragraph{Approximation of max-SC}To approximate max-SC, we can the existing algorithms for weighted max-SAT, which is a well-studied problem in the literature. We refer to the sequence of papers \cite{GoWi95,AsWi02,Asa05,PoSch11}. We form a SAT from our segment cover instance as follows. Any maximal sub-interval $J \subset I=[0,1]$ that does not contain an endpoint defines a clause, and in it are literals corresponding to uncertain segments covering the interval $J$. See the proof of Lemma \ref{l:equival}. We assign the length of $J$ as the weight of the corresponding clause. Given we have an algorithm for weighted max-SAT with approximation ratio $0<c'<1$, then clearly we have an algorithm with the same ratio for max-SC.

It is interesting to see these upper and/or lower bounds improved.

\small
\bibliographystyle{abbrv}

\bibliography{ref}

\end{document}